\documentclass[12pt]{article}
\usepackage{amsmath,amssymb}
\usepackage{graphicx}
\usepackage{amsmath}
\usepackage{float}
\newcommand {\be}{\begin{equation}}
 \newcommand {\ee}{\end{equation}}
 \newcommand {\bea}{\begin{array}}
 
 \newcommand {\eea}{\end{array}}

\textwidth=6.5in \hoffset=-.75in \voffset=-1in \numberwithin{equation}{section}
\numberwithin{figure}{section}

\def\e{{\epsilon}}

\def\l{\ell}

 \def\p{\partial}

\def\0{{(0)}}
\def\1{{(1)}}
\def\2{{(2)}}
 \def\cL{{\cal L}}

\def\<{\langle }
\def\>{\rangle }
\def\[{\left[}
\def\]{\right]}

\def\z{{\zeta}}

\begin{document}

\begin{titlepage}

\vskip1cm
\begin{center}
{~\\[140pt]{ \LARGE {\textsc{Kerr-Bolt Black Hole Entropy and Soft Hair }}}\\[-20pt]}
\vskip2cm

\end{center}
\begin{center}
{M. R. Setare \footnote{E-mail: rezakord@ipm.ir}\hspace{1mm} ,
A. Jalali \footnote{E-mail: alijalali.alijalali@gmail.com}\hspace{1.5mm} \\
{\small {\em  {Department of Science,\\
 Campus of Bijar, University of Kurdistan, Bijar, Iran\\
 Research Institute for Astronomy and Astrophysics of Maragha (RIAAM),
P.O. Box 55134-441, Maragha, Iran }}}}\\
\end{center}
\begin{abstract}
Recently it has been speculated that a set of infinitesimal  ${\rm Virasoro_{\,L}}\otimes{\rm Virasoro_{\,R}}$   diffeomorphisms exist which  act non-trivially on the horizon of some black holes such as kerr and Kerr-Newman black hole \cite{Haco:2018ske,Haco:2019ggi}. Using this symmetry in  covariant phase space  formalism one can obtains Virasoro charges as surface integrals on
the horizon. Kerr-Bolt spacetime is well-known for its asymptotically topology and has been studied widely in recent years. In this work we are interested to find conserved charge   associated to the Virosora symmetry of Kerr-Bolt geometry using   covariant phase space formalism. We will show right and left central charge are $c_R=c_L=12 J$  respectively. Our results also show good agreement with Kerr spacetime in the  limiting behavior.

\end{abstract}
\vspace{1cm}

Keywords: Black Holes, Space-Time Symmetries, Kerr-Bolt geometry

\end{titlepage}

\section{Introduction}
String theorists many years ago have  been found a non-abelian virasora symmetry which act non trivially on the black hole horizon \cite{Strominger:1996sh,Horowitz:1996fn}. This achievement leads to the black hole  microscopic entropy
of the black hole and reproduce the macroscopic area law  without reliance on stringy microphysics. In the context of proposed Kerr/CFT correspondence \cite{3}, the microscopic entropy of four-dimensional extremal Kerr black hole is obtained by studying
the dual chiral conformal field theory associated with the diffeomorphisms
of near horizon geometry of the Kerr black hole. Castro,
et. al \cite{Castro:2010fd} have given evidence that the physics of non-
extremal Kerr black holes might be captured by a conformal field theory. The authors have discussed that the wave equation for scalar fields in the Kerr spacetime has a certain conformal symmetry at low frequencies and in the region close to the horizon of the black hole, which they referred to as a hidden conformal symmetry. The initial computation of Castro et. al was repeated in various kind of black holes, and was shown to always lead to the same conclusions \cite{Chen:2010as,Wang:2010qv,Chen:2010bd,Setare:2010cj}. This kind of symmetries show
 left-right Virasoro pair on the horizon of black hole with $c_L=c_R=12J$ central charge.  Moreover, assuming the existence of a unitary Hilbert space this computation leads to the microscopic entropy of the dual CFT
by Cardy formula which possess perfect match to Bekenstein-Hawking entropy of black hole.

Recently diffeomorphisms and specially Infinite-dimensional symmetries
have become more attractive. The story began with Bondi and his et al works on gravitational waves \cite{  Bondi:1962px,Sachs:1962wk} and
 has been put on firmer footing with the discovery that relates the diffeomorphisms  to   soft graviton \cite{Weinberg:1965nx,He:2014laa,Zhang:2017geq}.

The effect of diffeomorphisms on the horizon of a four dimensional Kerr black   hole  was studied in \cite{Hawking:2016msc,Hawking:2016sgy} and introduced distinguishing features of black hole which named as soft hair.

More recently
in  \cite{Haco:2018ske,Haco:2019ggi}   authors were inspired by  this conformal symmetry and considered   general class of ${\rm Virasoro_{\,L}}\otimes{\rm Virasoro_{\,R}}$  diffeomorphisms of a generic spin
J Kerr black hole and Kerr-Newman black hole. Then they were going to use  the covariant formalism,  and  sought   existence of a well-defined central
charge $c_L=c_R=12 J$.

In the following work we are going to extend this paper to the Kerr-Bolt 4D spacetime.  We also find the left and right central charge as $c_L=c_R=12 J$.
Our work is organized in three section. First of all we briefly look at conformal symmetry of Kerr-Bolt space time. Then we continue with a short review  on symmetry in general relativity and  covariant phase space formalism to find out  conserved charge  in general relativity.
 In section three we present our result of Iyer-Wald charge and Wald and Zoupas counterterm. Finally we calculate entropy   of Kerr-Bolt black hole in the last section.
\section{Confromal coordinate and hidden conformal symmetry of Kerr-Bolt black hole}
The Kerr-Bolt spacetime with NUT charge $p$ and rotational parameter $a$, is given by the line element
\begin{eqnarray}
ds^{2}&=&-\frac{\Delta(r)}{\rho^{2}}[dt+(2p\cos(\theta)-a\sin^{2}(\theta)) d\phi]^{2}
+\frac{\sin^{2}(\theta)}{\rho^{2}}[a dt-(r^{2}+p^{2}+a^{2})d\phi]^{2}
\nonumber \\
&+& \frac{\rho^{2}}{\Delta(r)}+\rho^{2} (d\theta)^{2},
\label{intro1}
\end{eqnarray}
where
\begin{eqnarray}
\rho^{2}&=&r^{2}+(p+a\cos(\theta))^{2},\\
 \label{intro2}
\Delta(r)&=& r^2-2 M r +a^{2}-p^{2}.
\label{intro3}
\end{eqnarray}
The Kerr-Bolt spacetime (\ref{intro1}) is exact solution to Einstein equations. The inner $r_-$ and outer $r_+$ horizons of spacetime (\ref{intro1}) are the real roots of $\Delta(r)=0$, i.e.
\begin{equation}
r_+=M+\sqrt{-a^2+M^2+p^2}\qquad\qquad r_-=M-\sqrt{-a^2+M^2+p^2}
\end{equation}
This solution shows interesting  $SL(2,R)_L \times SL(2,R)_R$ virasoro symmetry.  To be more precise,  by considering massless scalar field $\Phi$ in  Kerr-Bolt background, i.e,
\begin{equation}
\Box\Phi=\frac{1}{\sqrt{-g}}\partial_{\mu}(g^{\mu\nu}\partial_{\nu})\Phi=0,
\label{intro4}
\end{equation}
and taking the expansion of scalar field as
\begin{equation}
\Phi(t,r,\theta,\phi)=\exp(-i m\phi+i\omega t) S(\theta) R(r).
\label{intro5}
\end{equation}
one can shows  that for the low frequency in the near region  or
\begin{eqnarray}
\omega M, \omega n &\ll& 1, \\
 r &\ll& \frac{1}{\omega},
\label{intro10}
\end{eqnarray}
the angular equation $S(\theta)$ significantly simplifies as
\begin{equation}
[\frac{1}{\sin(\theta)} \partial_{\theta}(\sin(\theta)\partial_{\theta})-\frac{m^2}{sin^2\theta}] S(\theta)=-l(l+1)S(\theta).\label{intro88}
\end{equation}
as well as radial component that takes the following form:
\begin{equation}
 \begin{array}{cc}
[\partial_{r}(\Delta\partial_{r})+\frac{-4Mmar\omega+m^{2}a^{2}-4n^{2}ma\omega}{\Delta(r)}+(\Delta+4(Mr+n^{2})-a+4n)
\omega^{2}]R(r)=l(l+1)R(r),
\end{array} \label{intro11}
\end{equation}
The virasoro symmetry could be seen manifestly  in the conformal coordinates $\omega^+, \omega^-$ and $y$ which are defined in terms of coordinates $t,r$ and $\phi$ as
 \cite{Ghezelbash:2010vt,Castro:2010fd}
\begin{eqnarray}
 \omega^{+}&=&\sqrt{\frac{r-r_{+}}{r-r_{-}}}\exp(2\pi T_{R}\phi+2n_{R}t),
  \label{intro14}\\
  \omega^{-}&=&\sqrt{\frac{r-r_{+}}{r-r_{-}}}\exp(2\pi T_{L}\phi+2n_{L}t),
  \label{intro15}\\
  y&=&\sqrt{\frac{r_{+}-r_{-}}{r-r_{-}}}\exp(\pi( T_{R}+T_{L})\phi+(n_{R}+n_{L})t),
\label{intro16}
 \end{eqnarray}
 where
\begin{equation}
 T_{R}=\frac{r_{+}-r_{-}}{4\pi a}, ~~~~~~T_{L}=\frac{r_{+}+r_{-}}{4\pi a}+\frac{p^{2}}{2\pi a M},
\label{intro17}
\end{equation}
and $n_{R}=0,n_{L}=-\frac{1}{4M}$.
In the conformal coordinate it is obvious that past horizon is  located in $\omega^-=0$,  $\omega^+=0$ is the future horizon and  the bifurcation surface $\Sigma_{\text{bif}}$ is obtained by $\omega^{\pm}=0$.
We also define the left and right moving vector fields
\begin{equation}
H_1=i\partial_{+},~~~~H_0=i(\omega^{+}\partial_{+}+\frac{1}{2}y
\partial_{y}),~~~~~H_{-1}=i((\omega^{+})^2\partial_{+}+\omega^{+}y\partial_{y}-y^2\partial_{-}),~~~~
\label{intro19}
\end{equation}
and
\begin{equation}
\bar{H}_1=i\partial_{-},~~~~\bar{H}_0=i(\omega^{-}\partial_{-}+\frac{1}{2}y\partial_{y}),
~~~~~\bar{H}_{-1}=i((\omega^{-})^2\partial_{-}+\omega^{-}y\partial_{y}-y^2\partial_{+}),
\label{intro20}
\end{equation}
respectively.
The vector fields (\ref{intro19}) satisfy the $SL(2,R)$ algebra
\begin{equation}
 ~~[H_0,H_{\pm1}]=\mp i H_{\pm 1},~~~~~~~~[H_{-1},H_1]=-2iH_0,~~
\label{intro21}
\end{equation}
and similarly for $\bar{H}_1,\bar{H}_0$ and $\bar{H}_{-1}$.

The quadratic Casimir operator, in terms of coordinates $\omega^+, \omega ^-$ and $y$, is given by
 \begin{equation}
H^2=-H_{0}^2+\frac{1}{2}(H_1H_{-1}+H_{-1}H_{1})=\frac{1}{4}
(y^2\partial_{y}^2-y\partial_{y})+y^2\partial_{+}\partial_{-}.
\label{intro28}
\end{equation}
In terms of coordinates $t,r,\theta,\phi$, the Casimir operator (\ref{intro28}) as well as the other Casimir operator $\widetilde{H}^2$  reduces to the radial equation (\ref{intro11}),
 \be \begin{array}{cc}
H^2R(r)=\widetilde{H}^2R(r)=l(l+1)R(r)~~~
\end{array} \label{intro29}\ee
To leading and subleading order around
the bifurcation surface, the Kerr-Bolt metric becomes
\begin{eqnarray}
\label{kbm}
 \begin{split} ds^2&= {4 \rho_+^2 \over y^2} d w^+ dw^-
+  {16 J^2 \sin^2\theta \over y^2 \rho_+^2} dy^2 +\rho_+^2 d\theta^2 \\[6pt]  &
- {2w^+ (8\pi J)^2 T_R(T_R+T_L) +p(64 \pi a T_RJ\cos\theta) \over y^3 \rho_+^2} dw^- dy \\[6pt]  &
+ {8 w^- \over y^3 \rho_+^2} \big(A+B\big) dw^+ dy \\[6pt] &
+ \cdots, \end{split}
\end{eqnarray}
where
\begin{eqnarray}
A&=&- (4\pi J)^2T_L(T_R+T_L) + (4 J^2 + 4\pi J a^2 (T_R+T_L)  + a^2 \rho_+^2) \sin^2\theta\nonumber \\
B&=&-4pa\cos\theta\big(m^2+p^2+1/2 \rho_+^2\big)
\end{eqnarray}
here  $\rho_+$  is given by
\begin{equation}
\rho_+^2=r_+^{2}+(p+a\cos(\theta))^{2}=
(m+\sqrt{-a^2+m^2+p^2})^{2}+(p+a\cos(\theta))^{2}\\
\end{equation}
Now consider a general vector field
 \begin{equation}
 \zeta(\e)= \e \p_++\frac{1}{2}\p_+\e y\p_y,
\end{equation}
where $\e$ is any function of $w^+$.
 One can easily verified that they  obey the Lie bracket algebra
 \begin{equation}
 \[ \z(\e),\z(\tilde \e )\]=\z(\e\p_+\tilde \e-\tilde \e\p_+ \e).\end{equation}
 A   complete set of  functions that is invariant  under $2\pi$ azimuthal rotations are \cite{Hawking:2016sgy}
 \begin{equation}
\label{csf}\e_n={2 \pi T_R}(w^+)^{1+{in \over 2 \pi T_R}}. \end{equation}
So
 \begin{equation}
 [\z_m,\z_n]=i(n-m)\z_{n+m}.
\end{equation}
  The zero mode is
 \begin{equation}
\label{rtj} \zeta_0=2\pi T_R( w^+\p_++\frac{1}{2} y\p_y )=-i2\pi T_R H_0
\end{equation}
 Similarly for the left movers we have:
   \begin{align}
\label{dsg} \bar \zeta_n&=\bar \e_n \p_-+\frac{1}{2}\p_-\bar \e_ny\p_y,\cr
\bar\e_n&={2 \pi T_L}(w^-)^{1+{in \over 2 \pi T_L}},
\end{align}
  with
   \begin{equation}
 \bar \z_0=-i2\pi T_L\bar{ H}_0
\end{equation}
 They also satisfy  Vir$_L$ algebra with zero central charge.
  \begin{equation}
 [\bar \z_m,\bar \z_n]=i(n-m)\bar \z_{n+m},
\end{equation}
and the two sets of vector fields commute with one another
  \begin{equation}
 [\z_m,\bar \z_n]=0.\end{equation}
 \section{Covariant phase space and covariant charges}
Notion of conserve quantity has got crucial role in investigating of physics theories. If theory  has $k^a$ killing field and $T_{ab}$ stands for energy-momentum tensor then one can take $J^a=T^a_{b}k^b$ as conserve quantity associated with $k^a$.

However, in general relativity which is the diffeomorphism covariant theory,
there is no notion of the local stress-energy tensor of the gravitational field, so
conserved quantities  can not be defined by the above procedures. One of the outstanding method for  conserve quantity definition among many others \cite{t,b,cjm,ds,as,g}  is covariant phase space \cite{Compere:2018aar,wz}. Consider a theory with lagrangian ${\bf L} $ and dynamical field $\phi$. Variation of lagrangian yields
\begin{equation}
\delta {\bf L} = {\bf E}(\phi) \delta \phi + d {\mbox{\boldmath
$\theta$}}(\phi, \delta \phi) .
\label{dL}
\end{equation}
The equations of motion are ${\bf E}=0$ and ${\mbox{\boldmath
$\theta$}}$ refers  to the boundary term or symplectic potential. The presympletic current is defined as
\begin{equation}
{\mbox{\boldmath $\omega$}} (\phi, \delta_1 \phi, \delta_2 \phi) =
\delta_1 {\mbox{\boldmath $\theta$}} (\phi,\delta_2
\phi)-\delta_2{\mbox{\boldmath $\theta$}} (\phi,\delta_1 \phi)
\label{omega}
\end{equation}
If $\phi$ Satisfies the equation of motion then and $\delta \phi_i$ satisfies the linearized equations of motion one can shows that $\omega$ is exact, i.e $d  \omega=0$ \cite{wz}. The Noether current  define as follow:
\begin{equation}
{\bf j} = {\mbox{\boldmath $\theta$}} (\phi, {\cal L}_\xi \phi) -
\xi \cdot {\bf L}=d {\bf Q} + \xi^a {\bf {C}}_a ,
\label{j}
\end{equation}
General form of $Q$ finds out in \cite{iw}. When equations of motion satisfy the  $C_a$ term appears as a constraints, i.e., $C_a=0$.
Combination of  \eqref{omega} and \eqref{j}  lead to the following equation for the ${\mbox{\boldmath $\omega$}} (\phi, \delta_1 \phi, \delta_2 \phi)$:
\begin{equation}
{\mbox{\boldmath $\omega$}} (\phi, \delta \phi, {\cal L}_\xi \phi) =
\xi^a \delta {\bf {C}}_a + d (\delta {\bf Q}) - d(\xi \cdot
{\mbox{\boldmath $\theta$}}) .
\label{dj2}
\end{equation}

Now consider a diffeomorphism covariant theory with a vector field $\xi^a$.  The $\delta Q_\xi$ known as conserved quantity associated to the $\xi^a$ vector field and defines as follow, \cite{wz}.
\begin{equation}
\delta {\cal Q}_\xi = \Omega_\Sigma(\phi, \delta \phi, {\cal L}_\xi \phi) =
\int_\Sigma
{\mbox{\boldmath $\omega$}}(\phi, \delta \phi, {\cal L}_\xi \phi)=
\int_{\partial\Sigma}
\left(\delta {\bf Q} - \xi \cdot
{\mbox{\boldmath $\theta$}}\right)
\label{H}
\end{equation}
where $\Sigma$ is a Cauchy hyper-surface in the theory and has been chosen so that  integral over $\Sigma$ converges for all $\phi$.  In this work $\partial\Sigma$ stands for bifurcation horizon surface. A conserved quantity for a theory with asymptotic symmetry also includes an extra term. As Wald and Zoupas have mentioned in their work \cite{wz} conserved quantity for a theory which has infinitesimal asymtotic symmetry vector $\xi^a$ is read as
\begin{equation}
\delta {\cal Q}_\xi = \int_{\partial \Sigma} [\delta {\bf Q} - \xi \cdot
{\mbox{\boldmath $\theta$}}] + \int_{\partial \Sigma} \xi \cdot
{\mbox{\boldmath $\Theta$}}
\label{hsurf2}
\end{equation}
where  $\Theta$ is   a symplectic potential for the
pullback of the
symplectic current form ${\mbox{\boldmath $\omega$}}$ to  the boundary. Above calculation for general relativity in four dimension is straightforward  and well-known \cite{wz,Chandrasekaran:2018aop}. The general relativity has
 the Einstein-Hilbert  Lagrangian four-form,
\begin{equation}
L = { \epsilon \over 16 \pi }  R.
\end{equation}
The  variation is  \cite{Hawking:2016msc}
\begin{equation}
\delta L = - {\epsilon \over 16 \pi }  G^{ab} h_{ab} + d\theta[h,g],
\end{equation}
where $\delta_h$ generates the variation $g_{ab} \to g_{ab}-h_{ab}$. The presymplectic potential $\theta$ is the three form
\begin{equation}
\theta[h,g] = *{1 \over 16 \pi } (\nabla^b h_{ab}\, - \nabla_a h )dx^a,
\end{equation}
where * is the Hodge dual star.  The conserved charge in general relativity has two parts and obtain by integration of  symplectic current $\omega$ as follow:
\begin{align}
\delta{\cal{Q}}(\zeta,h,g)&=\frac{1}{16\pi}\left(\delta{\cal{Q}}_{IW}(\zeta,h,g)+\delta{\cal{Q}}_X(\zeta,h,g)\right)\nonumber\\
&=\frac{1}{16\pi}\int_{\partial\Sigma}\ast F_{IW}+\frac{1}{16\pi}\int_{\partial\Sigma}\zeta\cdot(\ast X),
\end{align}
the first term is the Iyer-Wald charge, where integrand is given by
\begin{align}
\label{fab}
 \begin{split} F_{IWab} = \frac{1}{2}\nabla_a\zeta_bh
+\nabla_ah^c{}_b \zeta_c
+\nabla_c\zeta_a\ h^c{}_b
+\nabla_ch^c{}_a\ \zeta_b
-\nabla_ah\ \zeta_b - a \leftrightarrow b. \end{split}
\end{align}
and the surface of integration is the bifurcation surface. The second term in (3.10) is the Wald-Zoupas counterterm \cite{wz}, where $X$ is a  one-form constructed from the geometry of the spacetime.  As authors in \cite{Haco:2018ske} clarified X includes $\Omega_a$ one-form which shows kind of rotational velocity of the horizon and reads as
\begin{equation}
\label{counter}  X = 2dx^ah_a^{~b} \Omega_b\,.
\end{equation}
where
\begin{equation}
\Omega_a=q_a^c n^b\nabla_c \l_b\label{ox1}
\end{equation}
The vectors $\l^a$ and $n^a$ are normal to the future and past horizon respectively and have been chosen so that $\l\cdot n=-1$. We also consider $q_{ab}=g_{ab}+\l_an_b+n_a\l_b$ as the induced metric on the bifurcation surface \cite{Hawking:2016sgy}.

Assuming integrability, the conserved charge associated to the diffeomorphisms also form an algebra under Dirac braket:
\begin{equation}
 \{  {\cal{Q}}_n,  {\cal{Q}}_m\}=(m-n) {\cal{Q}}_{m+n}+K_{m,n},
\end{equation}
where the central extension is given by \cite{Compere:2018aar}
\begin{equation}
 \label{cterm} K_{m,n}=\delta {\cal{Q}}(\z_n,\cL_{\z_m}g;g).
\end{equation}
Also it has been shown \cite{Compere:2018aar} that the ce
ntral term K must be constant on the phase space and is given by
\begin{equation}
 K_{m,n}={c_i m^3\over 12}\delta_{m+n},
\end{equation}
where $c_i=c_L$ for the left mover conserved charge while $c_i=c_R$  for the right mover conserved charge.
\subsection{Left mover charge}
We begin with left mover conserved charge  on $\Sigma_{\text{bif}}$ by taking $\zeta = \bar{\zeta}_m$ and $\bar{h}^{ab} = \mathcal{L}_{\bar{\zeta}_n} g^{ab}$ and smooth limit $\omega^-\to 0$ on the past horizon. Because the structure of Kerr-Bolt geometry is similar to Kerr metric, our steps also are similar to \cite{Hawking:2016sgy} calculation.  The only nonzero component of $F^{ab}$ in \eqref{fab} is $F_{IW}^{+y}$ and it can be evaluated as
\begin{equation}
 F_{IW}^{+y} = - 4 \bar{h}^{y+}_m \bar{\zeta}_n \Gamma^+_{+y}=-4 g^{+-} \partial_-\bar{\zeta}^y_m \bar{\zeta}_n \Gamma^+_{+y}
\end{equation}
Using the Kerr-Bolt metric component in \eqref{kbm} and the below integral
\begin{equation}
 \lim_{w^+_0\to 0}
 \int_{w^+_0}^{w^+_0e^{4 \pi^2 T_R}} {dw^+ \over w^+} = 4 \pi^2 T_R,
\end{equation}
We find out ${\cal{Q}}_{IW}$ part of conserved charge as:
\begin{equation}
\label{liw}
K_{IW n,m}=\frac{1}{16\pi}\int_{\Sigma_{\text{bif}}}d\theta\, d\omega^-\epsilon_{\theta+-y}F_{IW}^{+y}=2J\frac{T_L-\alpha}{T_L+T_R-\alpha}m^3\delta_{m+n}\,,
\end{equation}
where $\alpha=\frac{p^2}{2\pi M a}$ is the dimensionless constant.
Determination of the Wald- Zoupas counterterm is required to fix $\l^a$ and $n^a$ vector field. Since integration area, i.e. $\Sigma_{\text{bif}}$ for the left mover conserved charge  approaches to the past horizon, we take these vector as
\begin{equation}
 \l \sim y^{{2 T_L-2\alpha \over T_R + T_L-\alpha}} \p_-, \quad\; n \sim y^{{2 T_R \over T_R + T_L-\alpha}} \p_+.
\end{equation}
In fact we have modified the definition of $\l^a$ and $n^a$, introduced in  \cite{Hawking:2016sgy} which are normal to the past and future horizon respectively. Substituting these definitions of $\l^a$ and $n^a$ in \eqref{ox1} and
some algebraic calculation lead to the following result for the Wald- Zoupas counterterm.
 \begin{equation}
 \label{xf}K_{X n,m}=J{T_R - T_L+\alpha \over T_L+T_R-\alpha}m^3 \delta_{n+m}.
\end{equation}
So interestingly, summation of the two terms in \eqref{liw} and \eqref{xf} yield to the left central charge as
$$c_L=12 J$$
\subsection{Right mover charge}
Right mover conserved charge is obtained in a similar manner but now we approach to the future horizon by  $\omega^+\to 0$ and
$\zeta ={\zeta}_m$ and ${h}^{ab} = \mathcal{L}_{{\zeta}_n} g^{ab}$ and smooth limit $\omega^-\to 0$ on the past horizon.
In this case the only nonzero component of $F^{ab}$ in \eqref{fab} is
\begin{equation}
F_{IW}^{-y}=-4 h_m^{y-} \zeta_n^y \Gamma^-_{y-}=-4g^{+-}\partial_+\zeta^y_m \zeta_n^y \Gamma^-_{y-}
\end{equation}
So the central part of conserved right mover charge is read as:
 \begin{equation}
 K_{IW n,m}={1 \over 16 \pi }\int_{\Sigma_{\text{bif}}}d\theta dw^+\e_{\theta+-y}F_{IW}^{-y} =-4\int_{\Sigma_{\text{bif}}}d\theta dw^+\e_{\theta+-y}
g^{+-}\partial_+\zeta^y_m \zeta_n^y \Gamma^-_{y-}\label{rkiw}
\end{equation}
After integration over future horizon one can find out:
 \begin{equation}
 K_{IW n,m}=2J\frac{T_R}{ T_L+T_R+\alpha}m^3 \delta_{n+m}.\label{iwr}
\end{equation}
The only quantity that leaves is Wald-Zoupas for the right mover conserve charge. Following same steps  which we used in the left mover case, we obtain Wald-Zoupas counterterm so that its combination with \eqref{iwr} leads to the $c_R=12 J$ right central charge. This counterterm is:
\begin{align}
K_{Xn,m}(\zeta,h,g)=\frac{1}{16\pi}\int_{\partial\Sigma}\zeta\cdot(\ast X)=
J\frac{T_L - T_R -\alpha}{ T_L+T_R-\alpha}m^3 \delta_{n+m}\label{xr}
\end{align}
Finally adding two terms in \eqref{iwr} and \eqref{xr} lead to the
$$c_R=12 J$$
\section{The area law and black hole entropy}
By using  of $c_L=c_R=12J$ as given above, the temperature formulae
 \eqref{intro17} and the Cardy formula
 \begin{equation}
S_{Cardy}={\pi^2 \over 3}(c_LT_L+c_RT_R),
\end{equation}
yields the Hawking-Bekenstein  area-entropy law for  Kerr-Bolt black hole.
 \begin{equation}
 S_{BH}=S_{Cardy}=2\pi M\left(r_++\frac{p^2}{M}\right)= {Area \over 4 }.
\end{equation}
\section{Conclusion}
In this study, we have used  hidden conformal symmetry of four-dimensional rotating spacetimes with NUT twist and covariant phase space formalism to obtain central charge associated to this symmetry.

Now we know that rotating spacetimes with a NUT twist shows Kerr/CFT correspondence \cite{Ghezelbash:2009gy}. The Kerr/CFT correspondence states that
the near-horizon states of an extremal four  dimensional black hole could
be identified with a certain chiral conformal field theory.  Chiral conformal symmetry in dual  field theory of Kerr geometry
provides the necessary ground
for the author in \cite{Ghezelbash:2009gy} to calculate the left and right central charge as $c_L=c_R=12J$.

In this paper we are going to reproduce this achievement by using of  covariant hase space formalism and  ${\rm Virasoro_{\,L}}\otimes{\rm Virasoro_{\,R}}$  diffeomorphisms symmetry of Kerr-Bolt black hole. We have found that $c_R=12J$ for the right mover or $\omega^+\to 0$ limit while we have seen $c_L=12J$ when we approach to the past horizon, i.e, $\omega^-\to0$. These results have good agreement with previous results.


\end{document}